\documentclass[letterpaper, 12pt]{article}

\usepackage{mathrsfs}
\usepackage[colorlinks=true,linkcolor=black,citecolor=black,urlcolor=black]{hyperref}

\usepackage{amsmath}
\usepackage{amssymb}
\usepackage{amsthm}
\usepackage{epigraph}
\usepackage{graphicx} 
\usepackage{url}
\usepackage{fancyhdr}
\usepackage{color}
\usepackage[czech,english]{babel}
\usepackage{indentfirst} % odsadi i prvni odstavec v textu
\usepackage[T1]{fontenc}
\usepackage{authblk}
\usepackage[utf8]{inputenc}

\usepackage[margin=1in]{geometry}
\usepackage{lipsum}

\usepackage{setspace}

\begin{document}
\begin{titlepage}
   \begin{center}
       \vspace*{1cm}

       \textbf{\Large The cosmological inflation inside the cyclic model of the universe}

       \vspace{0.5cm}
        
       \textbf{Kanabar Jay}

     {\itshape kanabarjaymgsi@gmail.com\\
        M.G. Science Institute, Gujarat University, Ahmedabad 380009, India\\
       }
\textbf{Maxim Khlopov}

     {\itshape khlopov@apc.in2p3.fr\\
        Virtual Institute of Astroparticle Physics, 75018 Paris, 
        France\\
       }

       \textbf{Jan Nov\'ak}

      {\itshape jan.novak@johnynewman.com (corresponding)\\
       Deparment of Physics, Faculty of Mechanical Engineering, Czech Technical University in Prague, Prague, 166 07, \\
       Czech republic}\\

   \end{center}

\begin{abstract}
Inflationary cosmology explains the homogeneity and large-scale structure of the universe through a brief epoch of accelerated expansion following the Big Bang. Cyclic cosmologies, in contrast, describe a universe undergoing successive phases of expansion and contraction and can generate primordial perturbations through alternative mechanisms while extending cosmic history beyond a singular beginning. Because these paradigms address different aspects of cosmic evolution, they are often treated separately. Here we explore the possibility that they may instead arise together. In a cosmological scenario involving two scalar fields, one field governs the cyclic evolution of the universe while the other drives an inflationary phase after each cosmological bounce, suggesting that inflation may be a recurring feature of a cyclic universe.\footnote{Essay written for the Gravity Research Foundation 2026 Awards for Essays on Gravitation}
\end{abstract}

\end{titlepage}

\selectlanguage{english}

%%%%%%%%%%
One of the central questions of science concerns the origin and evolution of the Universe itself. During the past century, advances in theoretical physics and observational astronomy have transformed cosmology from a largely philosophical endeavor into a precise quantitative science. The foundations of modern cosmology were established in the early twentieth century with the development of dynamical models of the Universe. In 1922, Alexander Friedmann discovered solutions of the Einstein field equations describing homogeneous and isotropic universes  \cite{friedmann1922}. 

These models allow for several possible cosmic histories: an open universe with negative curvature that expands forever, a closed universe with positive curvature that eventually recollapses, and a spatially flat universe that expands indefinitely with a gradually decreasing expansion rate. The physical relevance of these theoretical models became clear with the observational discovery by Edwin Hubble in 1929 that distant galaxies are receding from us, implying that the Universe is expanding \cite{hubble1929}. In all Friedmann-type cosmologies, this expansion traces back to an initial singular state, which later became known as the Big Bang, a term introduced in 1949 by Fred Hoyle \cite{hoyle1949}. The theoretical inevitability of such an initial state was further reinforced by the singularity theorems developed by Roger Penrose and Stephen Hawking, which showed under very general physical assumptions, these theorems imply that an expanding universe described by classical general relativity must originate from a spacetime singularity \cite{hawkingpenrose1970}.

Despite its successes, the standard Big Bang model leaves several deep puzzles unresolved. Among them are the remarkable homogeneity and spatial flatness of the universe and the origin of the tiny primordial perturbations that later grew into galaxies and large-scale structure. An influential proposal addressing these issues is the idea of cosmic inflation, a brief but intense period of accelerated expansion that smooths and flattens the universe while stretching microscopic quantum fluctuations to cosmological scales. These fluctuations become the seeds of structure, providing a compelling explanation for the nearly scale-invariant spectrum observed in the cosmic microwave background (CMB). Another important motivation for inflation arose from the monopole overproduction problem in grand unified theories, highlighted in the work \cite{zeldovich1978}, who noted that the existence of relic monopoles would strongly challenge the standard hot Big Bang picture. For a historical discussion of the development of inflationary cosmology and its early motivations, we refer the reader to the work \cite{liddlelyth2000,khlopov1999}. While answering puzzles of the Big Bang model, it leaves unanswered questions about the naturalness of the inflaton potential, the origin of initial conditions, and how inflationary dynamics are embedded in a full theory of Quantum Gravity (QG), \cite{baumann2022,mukhanov2005,weinberg2008,dodelson2003}.

Motivated by these questions, another line of thought has explored a radically different possibility, that the Big Bang was not the absolute beginning of cosmic history  but a bounce between successive cosmic cycles, allowing for recurring phases that shape each observable epoch. In cyclic cosmological models the universe undergoes a sequence of large-scale evolutionary phases, typically involving periods of expansion followed by contraction and a subsequent transition to a new expanding epoch. In these models, primordial perturbations are generated not by an exponential expansion, but by alternative processes depending on the cyclic model. However, constructing a physically consistent non-singular bounce or a smooth transition between two cycles is technically demanding, often requiring either exotic matter or non-perturbative quantum gravitational effects, and the cumulative entropy across cycles poses additional physical challenges.

Having their own conceptual issues, both paradigms aim to explain the early universe. They approach the same problem from different directions and are often viewed as competing alternatives. For these reasons,  we investigate a cosmological scenario in which cyclic evolution and inflationary dynamics may coexist. In particular, we consider a model involving two scalar fields, where one field governs the cyclic behavior of the Universe while the other drives an inflationary phase after each cosmological phase transition. Before developing this idea, it is useful to briefly review several influential cyclic cosmological scenarios that have been proposed in the literature. What distinguishes different cyclic models is how the universe returns to a very similar (or exact) previous state after a cyclic transition, particularly whether this process requires a period of contraction or not and if required, then how such contraction contributes to the dynamics. Understanding these models is essential for situating any proposal that combines cyclicity with inflationary dynamics.\\

One of the most intriguing cyclic scenarios proposed in recent decades is the idea of Conformal Cyclic Cosmology (CCC) developed by Roger Penrose \cite{penrose2006,penrose2010,penrose2012}. In this framework the universe is envisioned as a succession of cosmological epochs called aeons connected smoothly to the other next aeon via conformal transformations. The current universe represents just one aeon in this sequence, and each begins with a Big Bang-like state and evolves through the familiar stages of cosmic history before approaching a phase of accelerated expansion dominated by dark energy. As the universe in each aeon expands toward conformal infinity, its metric becomes increasingly stretched, causing the temperature and density to steadily decrease. Because of this property, the exponentially expanded and dilute future state of one aeon can be conformally rescaled into the extremely hot and dense state that characterizes the Big Bang of a new aeon. The transition between two such phases is known as a conformal crossover. An intriguing aspect of this picture is that events occurring in the late stages of a previous aeon might leave subtle observational signatures in the next aeon. 

CCC is mathematically complete within the framework of classical general relativity and conformal geometry. The model is elegant in its ability to connect cosmic history without invoking QG or exotic matter fields. Despite its conceptual elegance, CCC faces several important challenges. One commonly discussed issue concerns the requirement that all particle masses effectively vanish in the far future in order for the conformal mapping between aeons to be well defined. Another unresolved question concerns the thermodynamic evolution of entropy, particularly the enormous entropy associated with black holes and its relation to the low-entropy state expected at the beginning of a new aeon. Nevertheless, CCC provides an important example of a cyclic cosmological framework in which the end of one universe is directly connected to the birth of another.

Another influential approach to cyclic cosmology arises from the application of QG ideas to the early universe leading to bounce scenarios. In particular, Loop Quantum Cosmology (LQC) provides a framework in which quantum geometric effects modify the classical description of cosmological evolution derived from general relativity  \cite{bojowald2008}. LQC is based on the non-perturbative quantization program of gravity known as Loop Quantum Gravity (LQG). When the critical energy density determined by quantum geometric effects, approaches this critical value it implies that the Hubble parameter becomes zero and the contraction of the universe halts. Instead of encountering a singularity, the universe undergoes a transition from contraction to expansion known as a quantum bounce. This result has profound implications for cosmology. In the LQC framework, the Big Bang is replaced by a bounce connecting a prior contracting phase with the present expanding universe. In principle this mechanism allows the construction of cyclic cosmological scenarios in which successive periods of contraction and expansion occur. Furthermore, if scalar fields are present, the bounce can naturally set the initial conditions for a subsequent phase of inflation, providing a possible bridge between QG dynamics and the inflationary paradigm. Because the bounce occurs at Planckian energy densities, quantum gravitational effects may leave small but potentially observable imprints on the spectrum of fluctuations that later evolve into the large-scale structure of the universe. Although these corrections are typically predicted to be extremely small, future generations of cosmological observations may be capable of probing such signatures.

This approach connects the high-energy quantum effects with cosmological observables. Yet, LQC faces its own challenges. One of the problems it faces is the dynamics depend sensitively on the choice of quantization scheme, i.e., on LQG, which itself is not yet established as a complete theory of QG. Embedding additional scalar fields capable of driving a post-bounce inflationary phase remains technically nontrivial. Furthermore, while LQC resolves the singularity, it does not  address the details of entropy problem or the detailed transfer of cosmological information from one cycle to the next. Nevertheless, LQC demonstrates how quantum gravitational physics could resolve the classical Big Bang singularity and replace it with a dynamical bounce, thereby providing another possible realization of cyclic cosmic evolution.\\

A third class of cyclic scenarios arises from ideas tied to M-theory and was developed most prominently by Paul Steinhardt and Neil Turok \cite{steinhardt2002}. In this picture, our four-dimensional universe is realized on a brane that moves in a higher-dimensional bulk, cosmological evolution is driven by the relative motion and periodic collisions of parallel branes. Each collision appears, from the point of view of observers on a brane, as a Big Crunch followed immediately by a Big Bang. The ekpyrotic phase in this model involves ultra-slow contraction driven by a steep negative potential for the radion field, smoothing and flattening the universe by suppressing inhomogeneities and anisotropies. This mechanism addresses the same problems as inflation but through different dynamics. Perturbations are generated by converting entropy fluctuations into adiabatic curvature perturbations, producing nearly scale-invariant spectra. Observationally, this class predicts a blue tensor spectrum with index $n_T\approx 2$, yielding negligible gravitational wave amplitude on CMB scales. Thus, the ekpyrotic mechanism offers a distinct alternative to inflation, with unique signatures in perturbations and gravitational waves. 

While this model elegantly solves the problems of cyclic universe and entropy growth along with the problems for which we developed the theory of inflation, it also faces some challenges. One of the challenges is embedding the scenario consistently in a UV-complete theory requires control over higher-dimensional dynamics and moduli stabilization which is still an open area of research. The beauty of this model is a challenge to itself, since it explains the existing data, leading to absence of observational data that distinctly supports the ekpyrotic mechanism over the standard inflationary model. Nevertheless, the ekpyrotic/cyclic brane scenario is important because it demonstrates that very different microphysics and higher-dimensional considerations can solve many of the large-scale puzzles which are usually attributed to inflation while making distinctive predictions for observables, \cite{lehners2008}.\\ 

%One of the similar challenge faced by both the theory of inflation and cyclic cosmology models is that they invoke the idea of multiverse in context of eternal inflation and  cyclic theories of Ijjas and Steinhardt leading to the ideas of “cyclicmultiverse”. For a detailed account of this matter we recommend the reader to the following references.

The cyclic scenarios discussed above illustrate several distinct mechanisms through which the evolution of the universe might extend beyond a single Big Bang event. All these different approaches share a common feature that the present expanding universe may represent only one stage in a much longer cosmic history. At the same time, observational data provides strong evidence that a period of accelerated expansion, i.e., the cosmological inflation, might have occurred in the early universe. The remarkable agreement between theoretical predictions of inflationary models and measurements of the CMB suggests that some form of inflationary dynamics played a central role in shaping the large-scale structure of the universe.

However, these ideas need not necessarily be mutually exclusive. Rather than viewing cyclic cosmology and inflation as competing paradigms, one may consider whether both phenomena could occur within a single cosmological framework. A cyclic universe may in principle contain periods of accelerated expansion if the dynamical conditions following the transition from contraction to expansion allow for the temporary dominance of an inflationary component. 

A related possibility has been investigated in a different dynamical setting by Rossen Ivanov and Emil Prodanov \cite{ivanov2012}, who investigated a cyclic cosmological model in which inflationary phases arise from the thermodynamic behavior of a two-fluid system consisting of baryonic matter and a real gas quintessence component. In that approach the cyclic evolution emerges from the properties of a non-ideal equation of state, and the inflationary stages appear as part of the dynamical trajectories of the cosmological system in phase space. In their analysis they showed that this idea is possible for a wide range of initial data and a wide range of temperatures for real gas with an equation of state derived from the virial expansion and
for negative temperatures for a van der Waals real gas. Periodicity of the cyclic universe is lost too as the temperature of the Universe tends to zero. A natural question then arises that if scalar fields already play a central role in the physics of inflation, could a similar dynamical framework also accommodate the long-term cyclic evolution of the universe? 

The model considered here explores this possibility by embedding an inflation into a cosmological background whose large-scale evolution is governed by cyclic dynamics. This approach offers two important advantages. First, scalar fields provide the natural language in which most early-universe physics is currently formulated, including cosmological inflation and many proposals emerging from high-energy theory. Second, scalar field models allow a direct connection with the well-developed theory of cosmological perturbations. In this setting the curvature and entropy perturbations arise from the coupled dynamics of the scalar fields, and their spectra can be computed directly from the background solution of the system. Observable quantities such as the scalar spectral index and possible multi-field corrections are determined by geometric properties of the trajectory in field space, including the turn rate of the background evolution and the effective entropy mass. 

To investigate this idea we consider a model containing two scalar fields, denoted by $\phi$ and $\varphi$. The field $\phi$ governs the long-term cyclic evolution of the universe, while the field $\varphi$ is responsible for generating the inflationary stage shortly after the transition to expansion. The dynamics of the system can be described by the action 
\begin{gather}
S  = \int \sqrt{-g}\Big[\frac{R}{2} - \frac{1}{2}\partial_{\mu}\varphi\partial^{\mu}\varphi - \frac{1}{2}\partial_{\mu}\phi\partial^{\mu}\phi - \frac{1}{2}b(\varphi,\phi) \partial_{\mu}\varphi \partial^{\mu}\phi - V_{IC}(\varphi,\phi) - \beta^4(\phi)(\rho_M + \rho_R)\Big],  
\end{gather}
Here, $V_{IC}(\varphi,\phi)$ represents the potential governing the coupled dynamics of the two scalar fields, while the mixed term proportional to $b(\phi,\varphi)$ encodes a direct kinetic interaction between them, and the function $\beta(\phi)$ describes a coupling between the cyclic field and the matter-radiation sector. Such couplings naturally arise in cyclic cosmological models in which the field controlling the cycles is associated with an effective interbrane separation or another geometrical degree of freedom.% For a concise essay format, the derivations leading to the dynamical equations cannot be presented in full detail, since their complete development would require a much longer technical treatment. The following discussion therefore accompanies the equations with a conceptual interpretation of their structure and implications. 

When the action is varied with respect to the metric (spatially flat FLRW), one obtains the background cosmological equations that determine the evolution of the homogeneous universe. The Friedmann equations, which relate the Hubble expansion rate to the total energy density contained in the scalar fields and the matter-radiation components, take the following form in our model :
\begin{gather}
H^2  = \frac{1}{3}\Big{(}\frac{1}{2}\dot{\phi}^2 +\frac{1}{2}\dot{\varphi}^2 +\frac{1}{2}b(\phi,\varphi)\dot{\phi}\dot{\varphi} +  V_{IC}(\phi,\varphi) + \beta^4(\phi)(\rho_M+ \rho_R)\Big{)}
\end{gather}  
where $H = \dot{a}/a$ denotes the Hubble parameter and $a(t)$ is the cosmological scale factor describing the expansion of the universe. The first two terms represent the kinetic energies of the two scalar fields, and the other terms have the same meaning as described above. The second background equation describes the acceleration of the scale factor and determines whether the universe is undergoing expansion, contraction, or accelerated expansion. It can be written as:
\begin {gather}
\frac{\ddot{a}}{a} = - \frac{8\pi G}{3} (\dot{\phi}^2  + \dot{\varphi}^2 + b(\phi,\varphi)\dot{\varphi}\dot{\phi}-V_{IC} + \frac{1}{2} \beta^4\rho_M + \beta^4\rho_R).
\end{gather}
Both of these equations make clear how accelerated expansion can arise in the model. When the potential energy of the inflationary field dominates over the kinetic contributions, the universe experiences a phase of accelerated expansion. In the context of the present scenario, such a phase is interpreted as the inflationary stage that follows the transition from contraction to expansion within the cyclic cosmological evolution. The remaining two equations are obtained by varying the action with respect to the scalar fields themselves. These equations determine how the fields evolve in response to the cosmic expansion and to their mutual interactions. After performing the necessary variations and simplifying the resulting expressions, one arrives at the coupled system
\begin{gather}
(1- \frac{b^2}{4})\ddot{\varphi}  + 3H (1-\frac{b^2}{4})\dot{\varphi} + V_{IC,\varphi} - \frac{b}{2} V_{IC,\phi} =  \frac{1}{2} b \beta^3(\phi)\beta_{,\phi} \rho_M,  \nonumber\\
(1- \frac{b^2}{4})\ddot{\phi}  + 3H (1-\frac{b^2}{4})\dot{\phi} + V_{IC,\phi} - \frac{b}{2} V_{IC,\varphi} =  -\beta^3(\phi)\beta_{,\phi}\rho_M. 
\end{gather} 
These relations reveal several important features of the model. First, the evolution of the two scalar fields is explicitly coupled through both the kinetic interaction term and the derivatives of the potential. Second, the presence of the Hubble damping terms $3H \dot{\phi}$ and $3H \dot{\varphi}$ reflects the influence of cosmic expansion on the scalar fields. In particular, once the universe enters the expanding phase, these friction terms can slow the motion of the inflationary field sufficiently to place it in a slow-roll regime, thereby initiating a period of accelerated expansion.
The physical interpretation of this mechanism can be understood qualitatively as follows. During the contracting phase preceding the bounce or transition, the scalar fields evolve under conditions of rapidly changing curvature and energy density. Such dynamics can displace the inflationary field from the minimum of its potential. When the universe subsequently transitions into an expanding phase, the sign of the Hubble parameter becomes positive, and the resulting frictional effect suppresses the kinetic energy of the field. If the potential is sufficiently flat in the relevant region of field space, the field may then enter a slow-roll regime, producing a temporary period of cosmological inflation.

Within this framework, one can derive predictions for observable quantities characterizing the primordial perturbation spectrum.
Although the mathematical treatment of perturbations in this system can make use of techniques developed for multi-field inflation, the physical interpretation of the present model is fundamentally different. In conventional multi-field inflation, both scalar fields typically participate in generating a single inflationary phase in the early universe, which is not the case here since both scalar field are responsible for two different processes. The coupled dynamics of the scalar fields naturally give rise to both curvature and entropy perturbations, whose properties depend on the geometry of the trajectory followed by the system in field space. Quantities such as the turn rate of this trajectory and the effective mass of the entropy mode determine how fluctuations evolve and how they are converted into observable density perturbations.

 A particularly important result concerns the scalar spectral index, which determines the scale dependence of the primordial power spectrum measured in the CMB. For a wide range of inflationary potentials compatible with the present scenario, the leading prediction can be derived, which takes the familiar form
 \begin{equation}
n_s \simeq 1 - \frac{2}{N_{\rm CMB}},\ \ \ \mbox{where}\ N_{\rm CMB} \approx 50-60,
\end{equation}
where $N_{\rm CMB}$ denotes the number of e-foldings between the horizon exit of the cosmological modes and the end of the inflationary phase. In this way the model provides a direct link between the large-scale cyclic dynamics of the universe and the observable properties of primordial cosmological perturbations. %Although the detailed calculations leading to these results extend beyond the scope of this essay which can further reveal some interesting details and possible challenges, the structure of the equations above demonstrates how the coexistence of cyclic evolution and a transient inflationary epoch can be studied within a unified dynamical framework.
We finally conclude our discussion with one of the famous quotes of Werner Heisenberg \emph{"Not only is the Universe stranger than we think, it is stranger than we can think."},  thus encouraging exploration of the possibility that inflation may recur within an underlying cyclic universe, providing a framework in which cosmic structure emerges repeatedly across vast timescales. The purpose of this essay was to present and explore this conceptual possibility and to invoke new ways of thinking about the origin and continuity of the universe.

\end{document}